\documentclass[a4paper]{spie}  %>>> use for US letter paper
\usepackage{amsmath,amsfonts,amssymb}
\usepackage{graphicx}
\usepackage[colorlinks=true, allcolors=blue]{hyperref}

\title{High-uniformity TiN/Ti/TiN multilayers for the development of Microwave Kinetic Inductance Detectors}

\author[a,b]{Mario De Lucia}
\author[a,b]{E. Baldwin}
\author[a,b]{G. Ulbricht}
\author[a,b]{J. D. Piercy}
\author[a]{O. Creaner}
\author[c,a]{C. Bracken}
\author[a]{T. P. Ray}
\affil[a]{Dublin Institute for Advanced Studies, Astronomy and Astrophysics, 31 Fitzwilliam Place, D02XF86 Dublin, Ireland }
\affil[b]{School of Physics, Trinity College Dublin, College Green, Dublin 2, Ireland}
\affil[c]{Physics Deparment, Maynooth University, Maynooth, Kildare, Ireland}

\authorinfo{Further author information:\\ delucia@cp.dias.ie}

\begin{document} 
\maketitle

\begin{abstract}
Microwave Kinetic Inductance Detectors (MKIDs) are a class of superconducting cryogenic detectors that simultaneously exhibit energy resolution, time resolution and spatial resolution. The pixel yield of MKID arrays is usually a critical figure of merit in the characterisation of an MKIDs array. Currently, for MKIDs intended for the detection of optical and near-infrared photons, only the best arrays exhibit a pixel yield as high as 75-80\%. The uniformity of the superconducting film used for the fabrication of MKIDs arrays is often regarded as the main limiting factor to the pixel yield of an array. In this paper we will present data on the uniformity of the TiN/Ti/TiN multilayers deposited at the Tyndall National Institute and compare these results with a statistical model that evaluates how inhomogeneities  affect the pixel yield of an array.
\end{abstract}

\keywords{MKIDs, Titanium Nitride, Uniformity, Critical Temperature}

\section{INTRODUCTION}
\label{sec:intro}
Microwave Kinetic Inductance Detectors (MKIDs) are superconducting cryogenic detectors with many applications in multiple fields of science \cite{mkidsapplications}, notably visible and infrared astronomy\cite{MEC,DARKNESS,arcons,deshima-2020,toltec2018} and  particle physics \cite{CUORE2002,calder_colantoni_2018_2}. An MKID is a superconducting L-C circuit, often made of a meandered inductor and an interdigitated capacitor, that is capacitively coupled to a feed-line which is used to drive it and read it out \cite{Day2003}. When a photon strikes the MKID it breaks superconducting Cooper pairs and produces a number of quasi-particles. The number of which depends on the bandgap of the superconductor and the energy of the photon. The generated quasi-particles produce an increase in the kinetic inductance of the superconductor and consequently a shift in resonance frequency of the L-C circuit \cite{Day2003}. As the quasi-particles recombine into Cooper pairs the resonance frequency shifts back into its original position. The read out of an MKID occurs through an FPGA (Field Programmable Gate Array) board such as the ROACH (Reconfigurable Open Access Computer Hardware) board. The signal processing algorithm involves a Fast-Fourier Transform which channelises the signal into individual frequency bins in order for the electronics to process them. Although the resonant frequency of the MKID can be monitored and a change in it is associated with a detection event \cite{Day2003}, often a phase-monitoring is preferred. Since the number of quasi-particles generated depends on the photon energy, and the frequency (and phase) shift depends on the number of quasi-particles generated,  the change in resonance frequency is directly linked to the energy of the striking photon. \\

Due to their superconducting nature, Microwave Kinetic Inductance Detectors are single photon sensitive detectors in the UV, optical to near-IR range and are inherently energy-resolving detectors in this region of the electromagnetic spectrum. Being operated at low temperatures, around 100 mK, MKIDs exhibit no dark counts like CCDs would. Furthermore, multiple resonators can be coupled to the same feed-line provided that they resonate at unique resonance frequencies, in a scheme called Frequency Domain Multiplexing. As of 2022, up to 2000 resonators (evenly distributed across the 4-8 GHz octave) can be coupled to the same feed-line \cite{MEC}. \\ Current MKID arrays suffer from a limited pixel yield, only the best arrays exhibit 75-80\% of working resonators \cite{mazin2020}. One of the main limitations to the fabrication yield is non-uniformity  in thickness and stoichiometry of the superconducting thin films. Variations in these parameters can cause variations in the kinetic inductance of each individual resonator, resulting in uncontrolled resonance frequencies and potentially overlapping resonators.  
\section{MATERIAL CHOICE}
The material choice is of paramount importance, as the intrinsic properties of the superconductor define the performance of the MKID array.\\ 
The ideal superconductor would have the following characteristics: 
\begin{enumerate}
    \item Lowest critical temperature possible
    \item Low internal losses, therefore high $Q_i$
    \item High sheet inductance
    \item Easy to deposit 
    \item High uniformity and homogeneity
    \item Low reflectivity in the wavelength range of interest
\end{enumerate}
A low critical temperature means that upon striking, a photon of a given wavelength excites a higher number of quasi-particles, which is reflected in  a higher signal and a higher signal-to-noise ratio. Unfortunately, practical limitations come into place: the finite cooling power of the cryostat sets a limit to the critical temperature of the superconductor that can be used. In order to fully deplete the superconductor of un-paired electrons an MKID array should be operated at a temperature 8-10 times lower than the superconducting critical temperature\cite{Zmuidzinas} . Effectively, a cryostat optimised for an operating temperature of 100 mK automatically sets a lower limit for the critical temperature of the films at about 800 - 1000 mK.\\
Low internal losses, or equivalently a high value of $Q_i$ (internal quality factor), not only is reflected in a higher signal to noise ratio, but also represents a more responsive pixel. The larger the sheet inductance of the film, the smaller the possible pixel size. Minimising the pixel size is crucial to produce large arrays.\\
A highly uniform and homogeneous film is highly beneficial. Variations in thickness or elemental composition will affect the inherent properties of the superconductor: critical temperature and sheet inductance. Uncontrolled variations of these properties can result in degraded performance of the detector array. In particular, a local variation in kinetic inductance can give rise to uncontrolled shifts in the resonance frequency of the affected resonators. The resonance frequencies can thus overlap and effectively decrease the pixel yield of the array \cite{DeLucia-multiplexing,Paul-thesis}.
\newline
Ideally an elemental superconductor would be the preferred choice. They're usually rather easy to deposit and, being elemental, they would have no problem arising from inhomogeneity in stoichiometry. Among the options, aluminium (Al) stands out. Al is very often used in the MKIDs community \cite{al1,Mazin,mkidsapplications} for applications in the sub-millimetre and millimetre despite its low kinetic inductance. Furthermore, Al exhibits high reflectivity in the visible and near-infrared region of the electromagnetic spectrum. Other options are listed in table \ref{table-elementalsuperconductors}.
\begin{table}[h!]
    \centering
    \begin{tabular}{|c|c|c|c|}
    \hline      
         Element & T$_c$ (bulk) & Pros & Cons \\
    \hline
        Al & 1.2 K & well understood & Low L$_k$, high reflectivity\\
    \hline
        Ti & 0.4 K & less reflective than Al & low T$_c$, low L$_k$ \\
    \hline 
        Hf & 0.165 K & Already used for MKIDs & low T$_c$ \\
    \hline

    \end{tabular}
    \caption{ List of elemental superconductors with their critical temperature. Advantages and drawbacks of each elemental superconductor are also listed.}
    \label{table-elementalsuperconductors}
\end{table}
\newpage
While elemental superconductors are expected to be highly uniform, the identification of the ideal superconductor  is not trivial. One approach that has been successfully tried involves the deposition of superconducting films that have controllable critical temperature. One particular option is sub-stoichiometric Titanium Nitride (TiN$_x$). The critical temperature of bulk stoichiometric TiN (5.6 K \cite{TiN-tc}) is too high for the purpose of this application, but it is possible to achieve control over the critical temperature of through control over the stoichiometry of the film \cite{Vissers_2013}. However, the control of critical temperature is challenging and the films are prone to exhibit inhomogeneities across the wafer \cite{Vissers_2013,x,y} due to small variations of nitrogen content \cite{Vissers_2013}, further enhanced by technological limitations including gas-flow control during magnetron sputtering deposition.\\
One other option that has been explored in the last few years exploits multi-layered stacks of TiN and Ti. In this configuration, it is not the stoichiometry that controls the critical temperature but the relative thickness of the layers. If the layers are sufficiently thin, the quantum mechanical proximity effect \cite{vissers-TiNTiTiN,Tinkham} arises as the macroscopic coherence length of the superconductor is comparable to the thickness of the layers.  This produces a film that has a critical temperature which depends on the relative thickness of the three layers. This approach is promising as it involves layers of stoichiometric TiN  and Ti which can be deposited with increased control over their uniformity and homogeneity compared to sub-stoichiometric TiN$_x$\cite{vissers-TiNTiTiN}.
\section{DEPOSITION}
\begin{figure}[h!]
    \centering
    \includegraphics[width=0.75\textwidth]{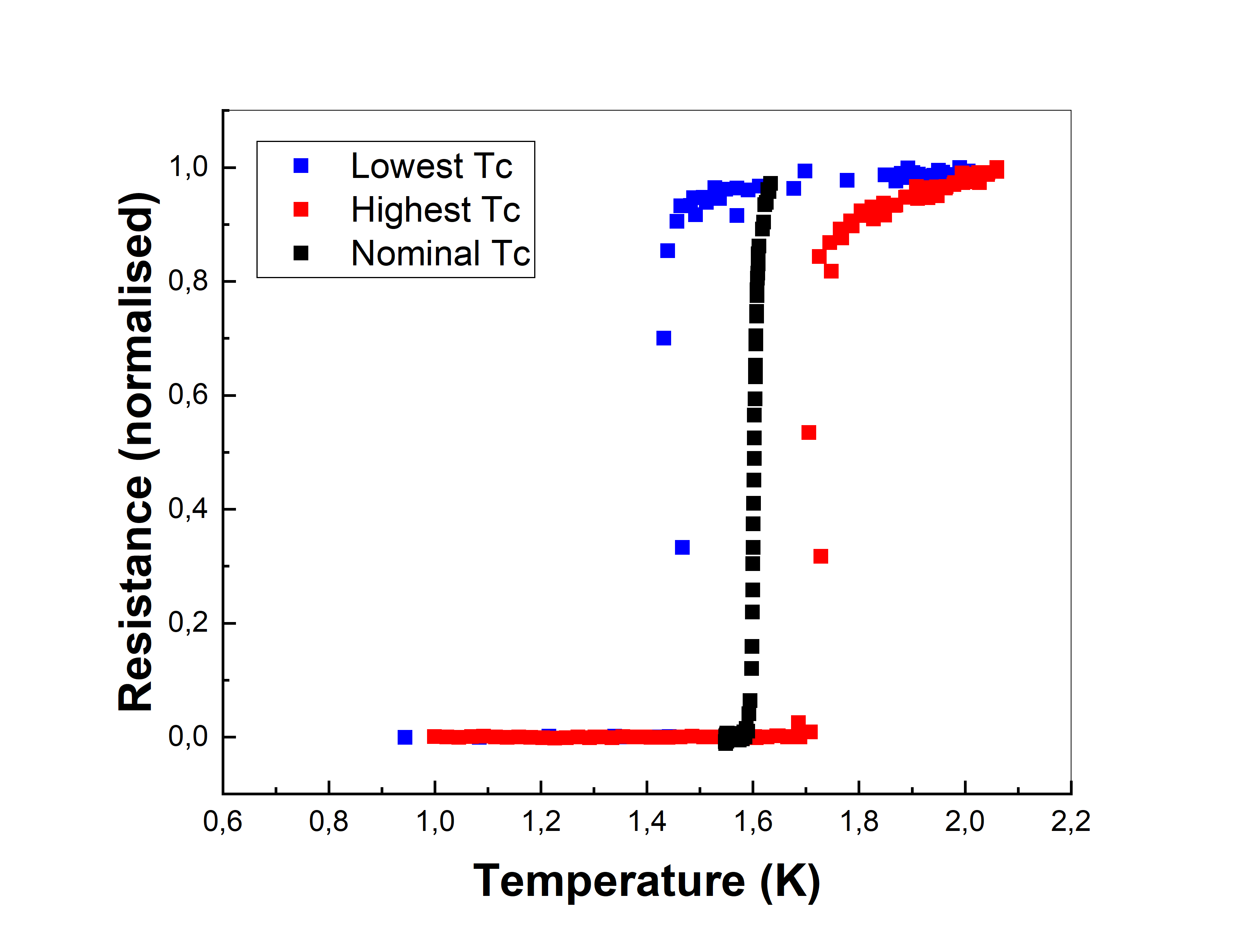}
    \caption{Critical temperature measurements of three samples: lowest T$_c$ (blue), highest T$_c$ (red) and nominal T$_c$ (black). These curves represent the reproducibility of our deposition process.}
    \label{TC-multiple}
\end{figure}

All the depositions of the superconducting layers were carried out at the Tyndall National Institute in Cork, Ireland. The Tyndall National Institute is a semi-commercial facility which is part of University College Cork and it currently has over 200 commercial partners.\\
\newline
The deposition of the layers occurred in a Nordiko 2500 magnetron sputtering system 10$^{-8}$ mbar base pressures. The substrate is a high resistivity ( $\rho > 10\, k\Omega\cdot cm$) (1,0,0) Si substrate. Before sputtering, the wafer undergoes a 60s  HF etch of the native silicon oxide (SiO$_2$) in order to avoid the detrimental effects of the amorphous SiO$_2$ to the performance of the detectors. The substrate is loaded in the deposition chamber within 10 minutes of the completion of the HF etch in order to prevent re-oxidisation. In the deposition chamber the sample is heated to a temperature of 150$^{\circ}$ C. The deposition of TiN occurs through reactive sputtering in an N$_2$ atmosphere whereas all the Ti atoms are sputtered off a Ti target.\\
\begin{figure}[h!]
\begin{minipage}[b]{0.5\textwidth}
    \includegraphics[width=0.9\textwidth]{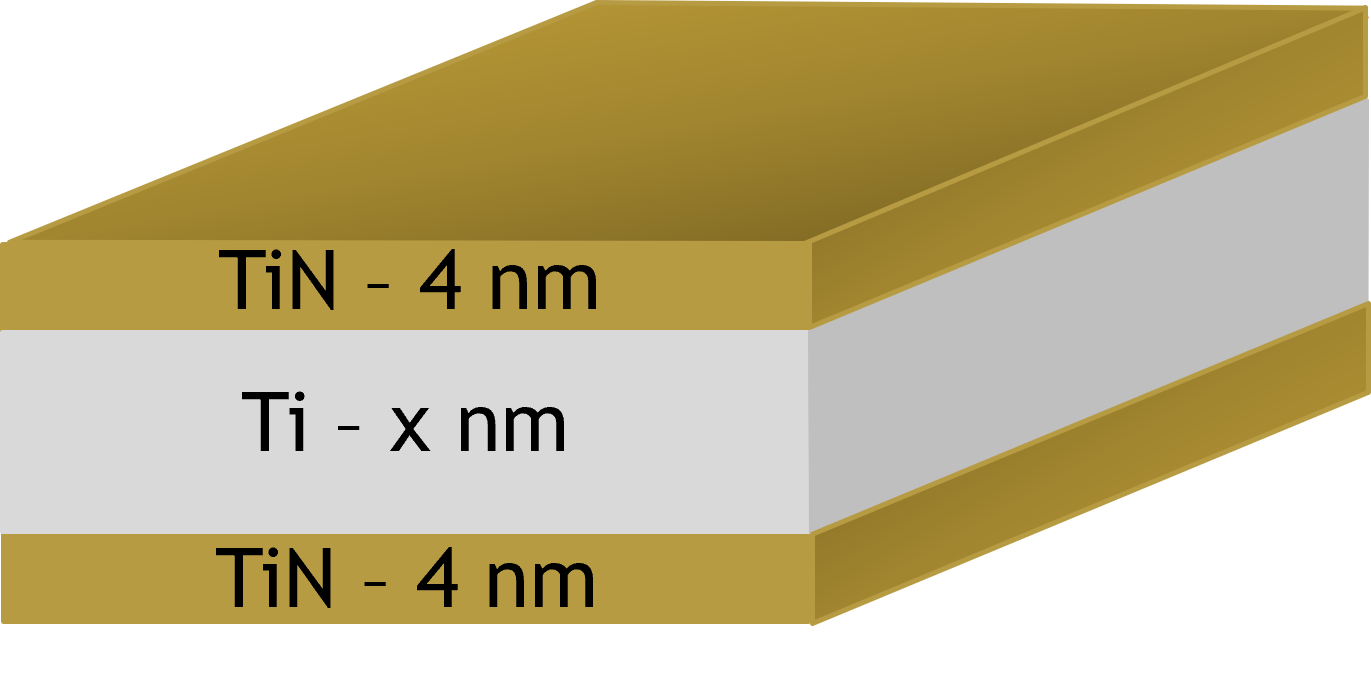}
\end{minipage}
\begin{minipage}[b]{0.5\textwidth}
  \includegraphics[width=0.9\textwidth]{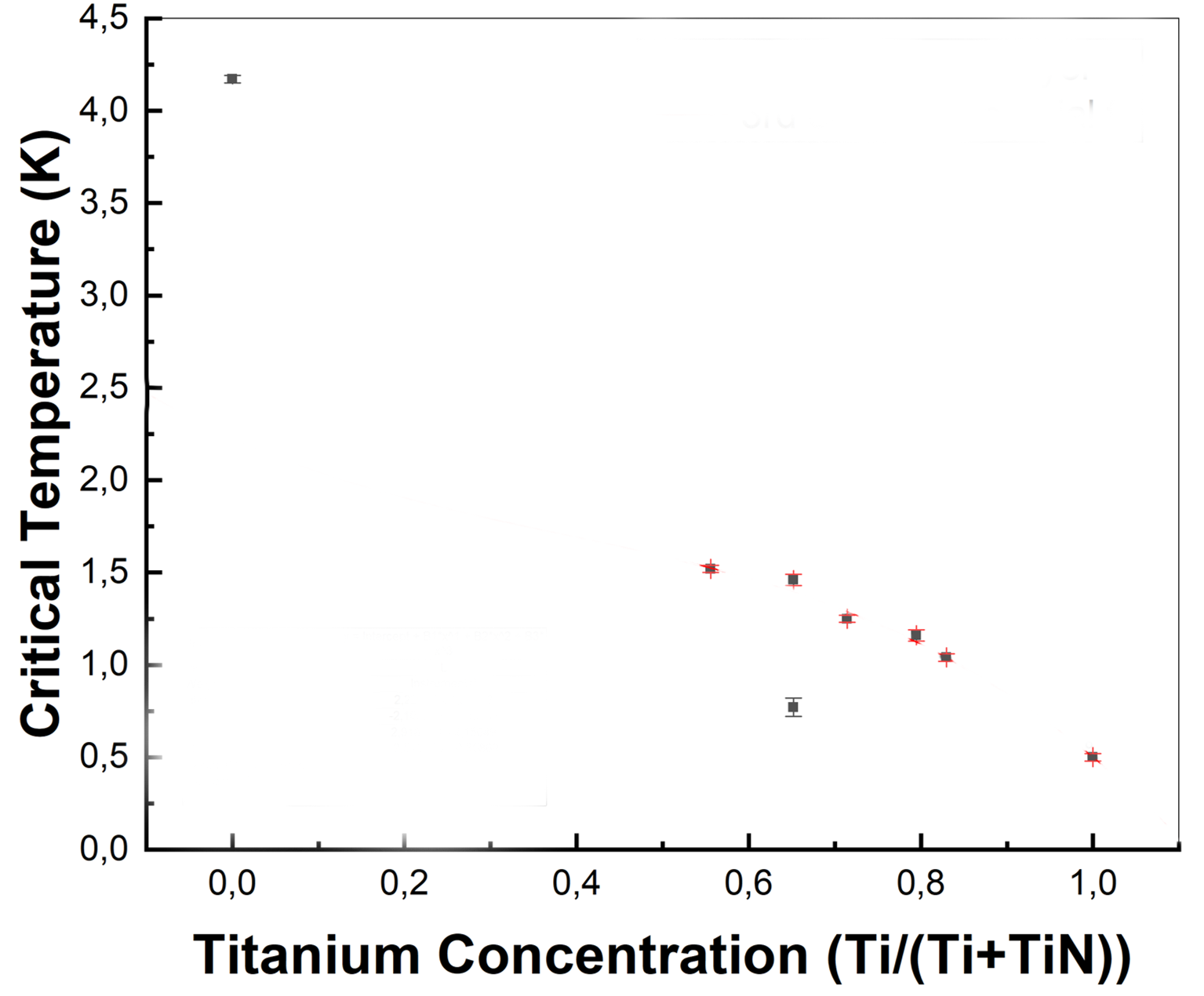}
\end{minipage}
  \caption{Left: schematic of one of our multilayers, the Ti which does not have a fixed thickness is sandwiched between two TiN layers each 4 nm thick. Right: critical temperature of the film as a function of the Ti to TiN ratio.}
  \label{fig:TiNstack} 
\end{figure}{}
\newline
All the films reported here have been deposited so that both the top and bottom TiN layers are nominally 4 nm thick as described in Figure \ref{fig:TiNstack} (Left), while the Ti layer varies. Figure \ref{fig:TiNstack} (Right) shows that by changing the thickness of the sandwiched Ti layer, it is possible to accurately control the critical temperature of the multilayered film continuously between the critical temperature of pure Ti and that of stoichiometric TiN.\\
It proved impossible to distinguish the Ti from the TiN layers through Scanning Electron Microscopy (SEM), however control over the thickness of the individual layers can be inferred through the characterisation of the critical temperature of the wafers. Information on the reproducibility of the critical temperature of the films can be indirectly regarded as information on the control and reproducibility of the thickness of each individual layer. Experiments carried out on wafers with nominal thickness 4 nm, 10 nm and 4 nm for the bottom TiN, the Ti and the top TiN layers respectively, showed good reproducibility within 100 mK of the target critical temperature (1.60 K).  Figure \ref{TC-multiple} shows the critical temperature of three samples from two different wafers measured. The three curves represent the range in which the critical temperature of a film with such  nominal thickness values would reasonably lie. The red curve represents the superconducting at the highest critical temperature recorded, the blue one is for the coldest and the black one indicates the behaviour of the nominal film.
\section{ SINGLE WAFER UNIFORMITY}
We next discuss how local and global inhomogeneities can affect the performance of an MKIDs array, in particular how a non-homogeneous film might account for an array with reduced pixel yield. \\ 
The sheet inductance of the thin film is a critical parameter for MKIDs. All of the resonators on the array are designed based on the assumption that the sheet inductance is constant across the full size of the chip. A variation in the sheet inductance of the superconductor results in a unpredictable variation in resonance frequency according to $f_0 = \frac{1}{\sqrt(LC)}$, and $L = L_g + L_k$. Here, L$_g$ represents the geometric inductance of the physical circuit and L$_k$ represents the kinetic inductance, which is an inherent property of the superconducting resonator and depends on the sheet inductance of the film. This variation could take the form of an abrupt local variation or a slow radial gradient across the diameter of the wafer. \\
The sheet inductance of a superconductor can be described through a simple approach that follows the discussion found in Annunziata et al. \cite{Annunziata} and Thinkam's \textit{Introduction To Superconductivity} \cite{Tinkham}. Given a superconducting element of length $l$, width $w$, and thickness $h$ cooled down to a temperature $T < T_{c0}$ and taking $k_B$ as Boltzmann's constant, Equation \ref{eq:Lk} shows how the sheet inductance of a superconducting element depends on its dimensions, in particular its thickness, and its normal-state sheet resistance $R_s$ as well as its critical temperature $T_{c0}$.

\begin{equation}
    L_{k}=\frac {l}{wh}\frac {R_{s}}{2\pi ^{2} 1.76 T_{c_0} }\tanh \left( \frac {1.76 T_{c_0} }{k_{B}T}\right)
\label{eq:Lk}
\end{equation}{}

\begin{figure}[h!]
    \centering
    \includegraphics[width=0.5\textwidth]{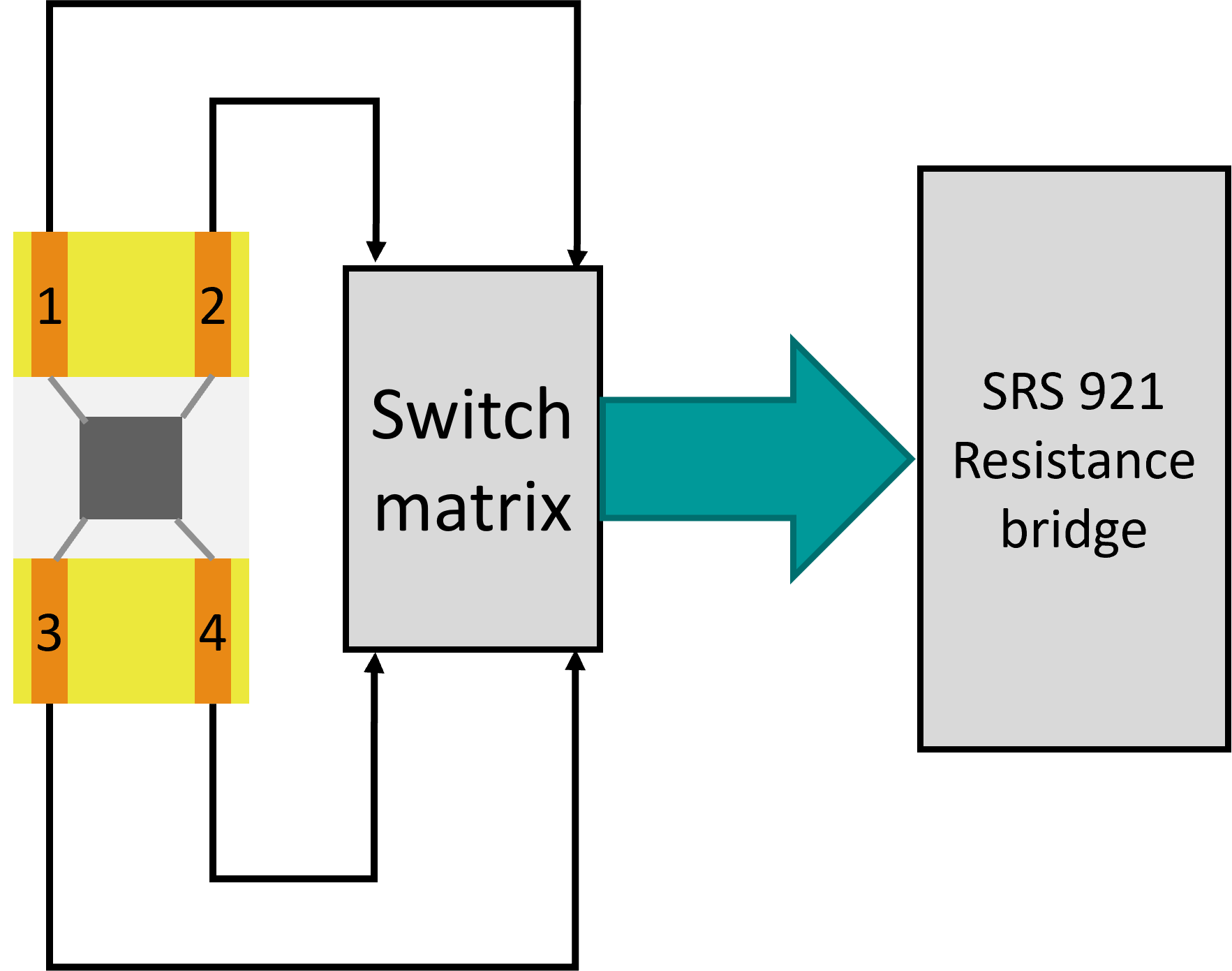}
    \caption{Experimental setup for the sheet resistance measurements. We have adapted some old printed circuit boards (in yellow) which had four copper pads (1,2,3,4). A glass slide for a microscope (light grey) was glued on the PCB in order to minimise charge transfer through the silicon substrate. The sample (dark grey) was glued on the slab and bonded to the four copper pads through a wire bonder. The circuit board is then connected to a switch matrix in order to perform 4 different four-point resistance measurements without having to re-bond. The resistance bridge is a SRS 921 from Stanford Research.}
    \label{fig:exp_setup}
\end{figure}

\begin{figure}[h!]
\begin{minipage}[b]{0.5\textwidth}
    \includegraphics[width=0.945\textwidth]{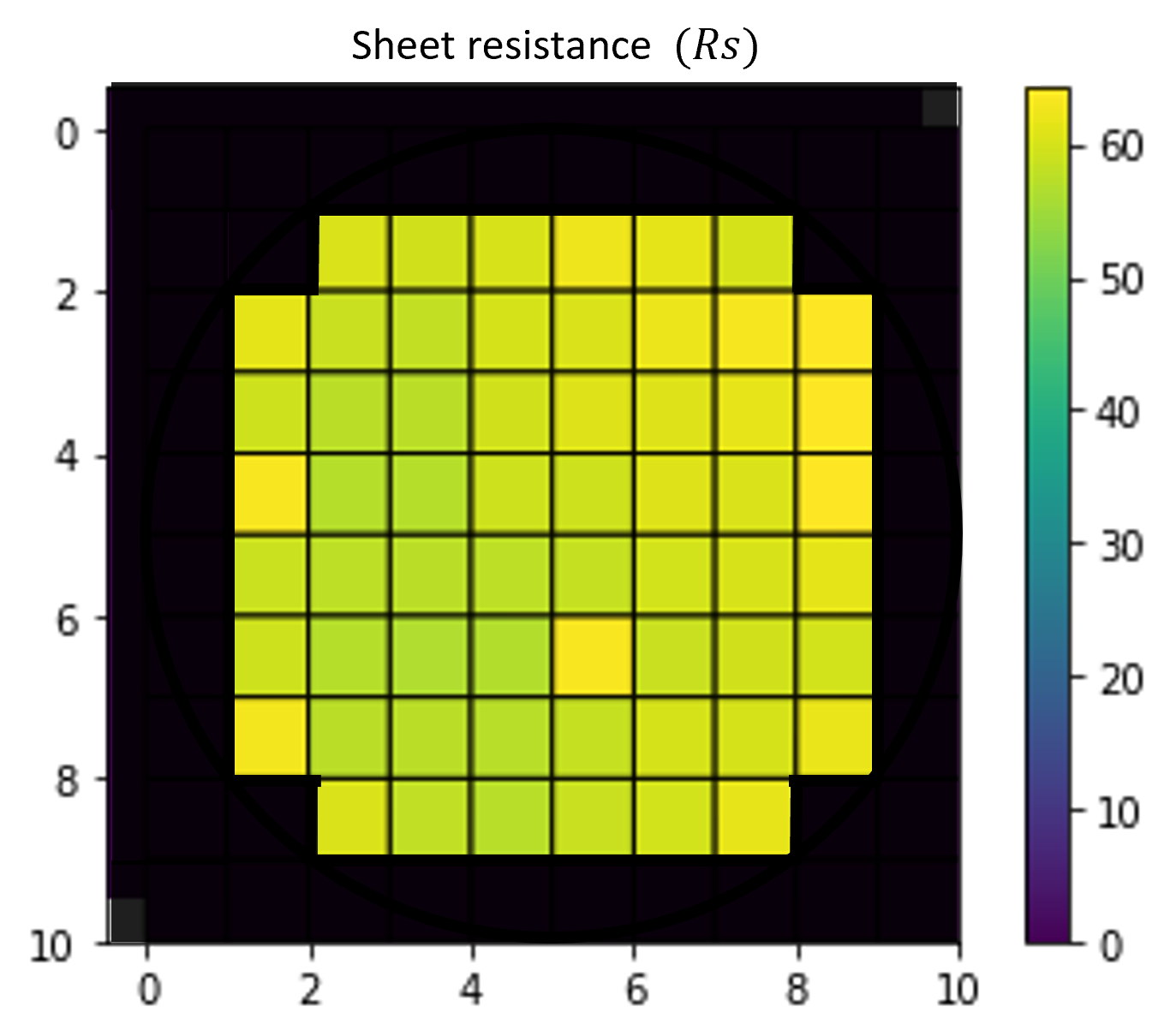}
\end{minipage}
\begin{minipage}[b]{0.5\textwidth}
  \includegraphics[width=\textwidth]{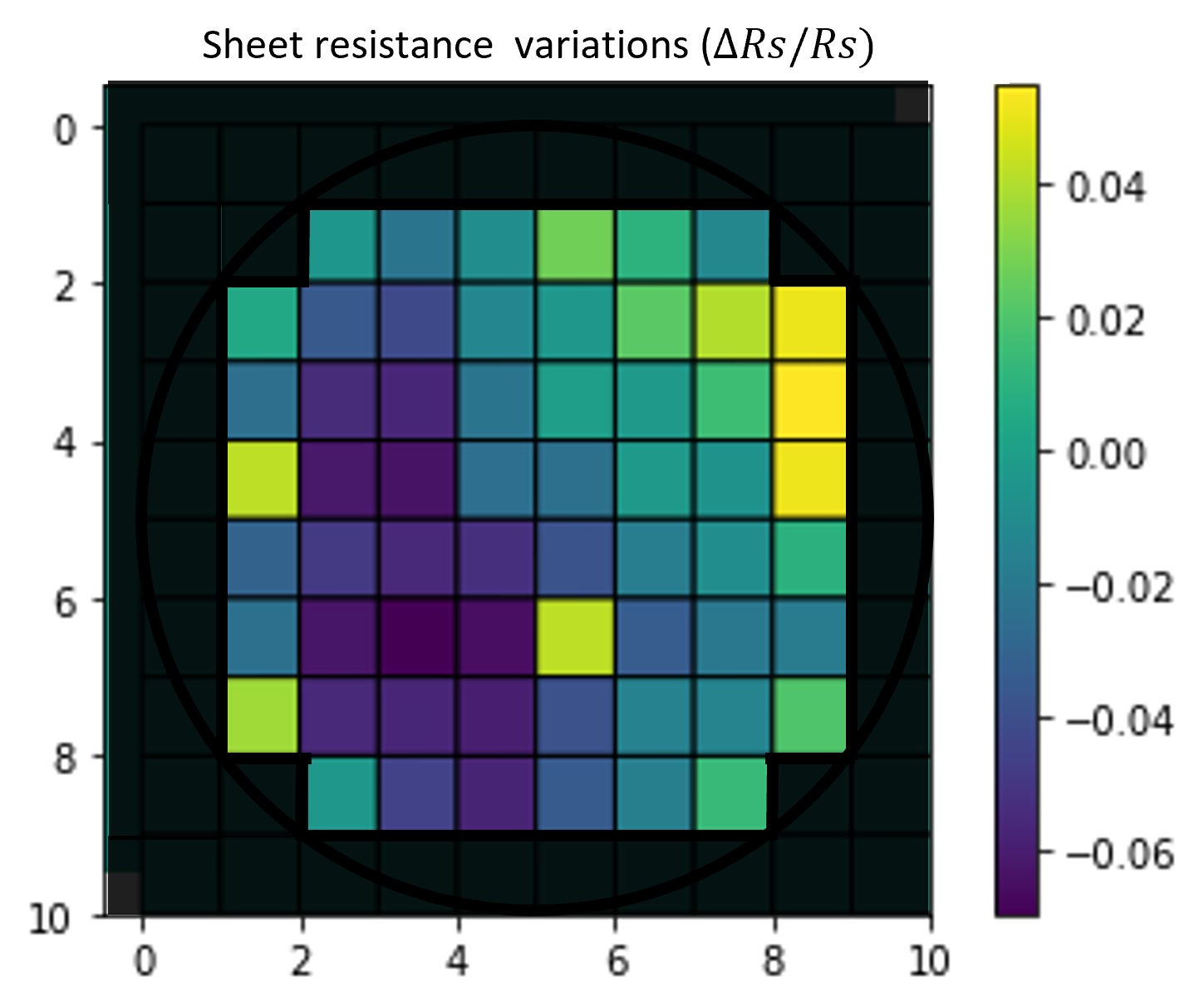}
\end{minipage}
  \caption{Left: sheet resistance map of the wafer produced through Van Der Pauw measurements. The grid is 1 cm $\times$ 1 cm . Right: same as left, but the map represents the variation of sheet resistance from average, in percentage.}
  \label{fig:resmap} 
\end{figure}

\begin{figure}[h!]
\begin{minipage}[b]{0.5\textwidth}
    \includegraphics[width=\textwidth]{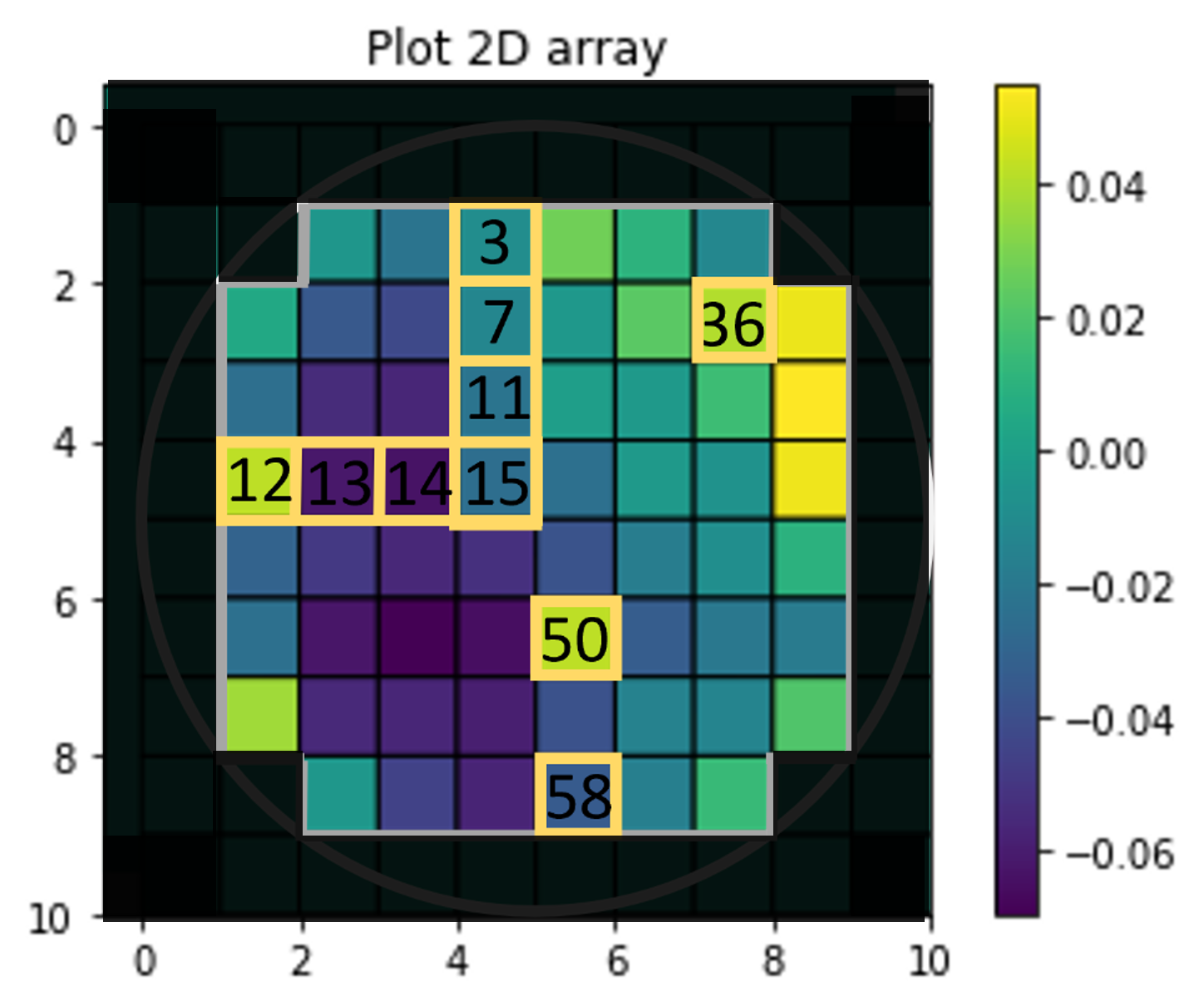}
\end{minipage}
\begin{minipage}[b]{0.5\textwidth}
  \includegraphics[width=1.1\textwidth]{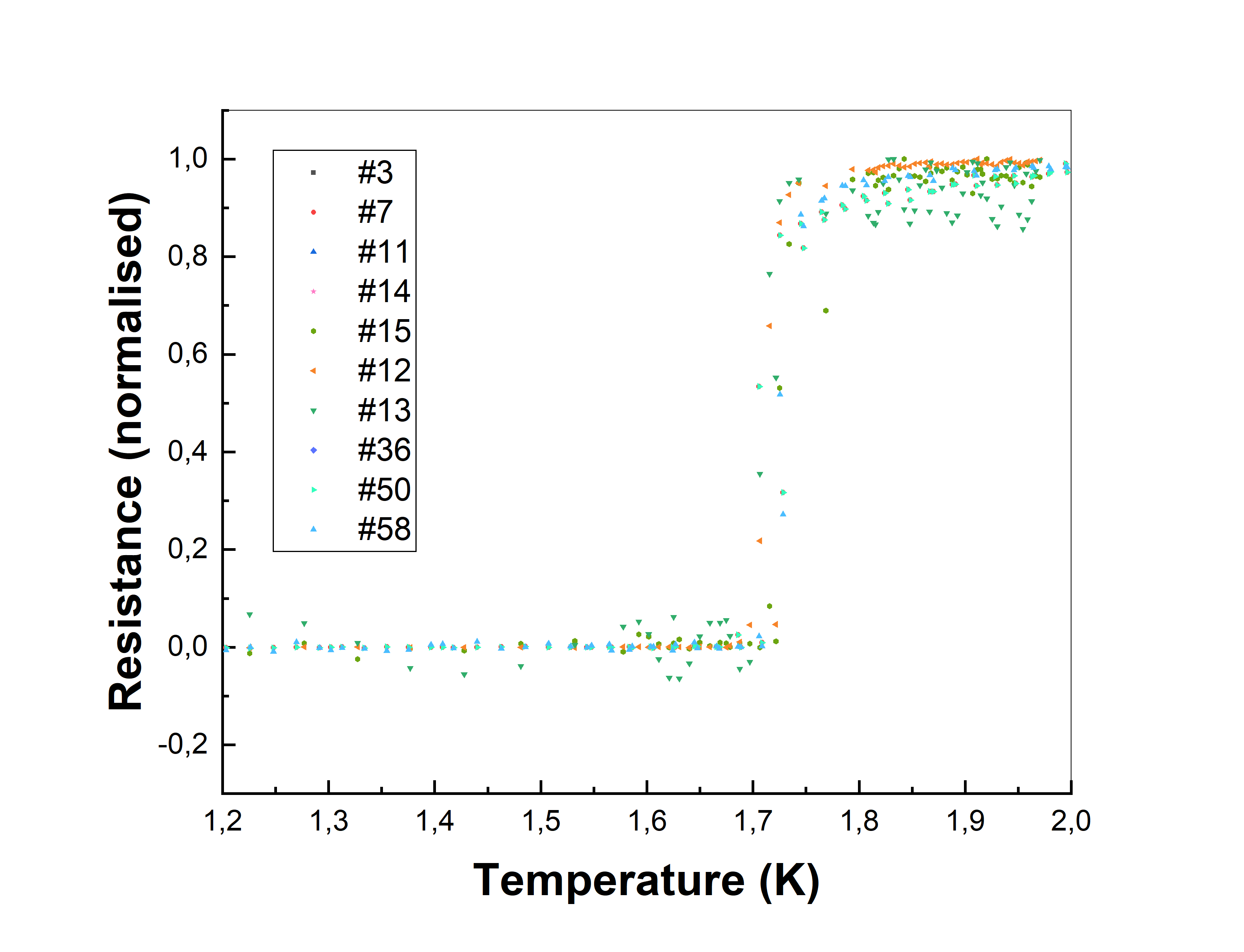}
\end{minipage}
  \caption{Left: map that identifies the samples for which the critical temperature was measured. Right: critical temperature measurements performed on the selected samples, as defined in the left panel of this figure.}
  \label{fig:selection} 
\end{figure}

Under the assumption that the critical temperature of the film does not vary too much across a 4 inch wafer, an assumption that we will further discuss later, we investigated how uniform the resistivity is across the wafer. This was achieved through 4-point Van Der Pauw measurements. The experimental setup is shown in Figure \ref{fig:exp_setup}. The sample was wire bonded to the Van Der Pauw setup and the resistance measurements, performed in the dark, were performed with a SRS 921 resistance bridge.\\
\\
The sample preparation was rather simple. A 4 inch wafer was diced according to a 1 cm $\times$ 1 cm square grid. Of all the samples thus produced, only the 60 inner ones, i.e. the ones with four straight edges, were taken into consideration for Van Der Pauw measurements. Van Der Pauw method correlates the resistance of the sample under examination with the sheet resistance of the thin film which appears in Equation \ref{eq:Lk}.\\
This way, we produced the sheet resistance map in Figure \ref{fig:resmap} (Left) and a map which shows the variation of the sheet resistance of each individual square, from the average value (Right). \\ 
{}\\
The map shows an average sheet resistance of $59.6\, \Omega/\square$ with variation across the wafer that are smaller than $\pm 6\%$. The accuracy of our measurements is within $\pm0.5\,\Omega/\square$.\\
In order to justify the previous assumption that the critical temperature of the film is rather uniform across the 4 Inch wafer, we have identified 10 square samples of interest (Figure \ref{fig:selection} (left): seven across two perpendicular radii (3,7,11,12,13,14,15) and three further squares that were regarded as particularly interesting because of large variations in resistivity as per the previous Van Der Pauw measurements (36,50,58). All ten chiplets exhibited a critical temperature of 1.70 K $\pm$ 0.01 mK (Figure \ref{fig:selection} (Right)), which validates our previous assumption and suggests that the relative thicknesses of the 3 layers do not vary significantly across the 4'' wafer. This will be further investigated through SEM scans. \\
{}\\
\section{MODELLING}
\subsection{Fabrication Yield}

 The fabrication of the detectors for such instruments is most often performed in research-grade clean rooms, through processes which are almost completely human-controlled and not automated. These processes have plenty of room for optimisation, although the fraction of non-responding pixels is less than $1\,\%$.
 \\
 Two or more resonators that resonate too near in frequency when they were not designed for this purpose, are said to be \textit{clashing} or \textit{colliding}. It is expected that random variations in the normal state sheet resistance of the superconductor (R$_s$) can have a detrimental impact on the fabrication yield. 
 \\
 The current state of the art is that most arrays exhibit a fabrication yield of $75\,\%$ and only for the best ones it goes as high as $80\,\%$ \cite{mazin2020}.\\
The model here discussed is based off the approach described by Liu et al.\cite{yield_liu}. 
This simple model relies on a few basic assumptions:
\begin{enumerate}
    \item the array of resonators is infinite on both sides of resonator $\#0$  
    \item the distance, in frequency space, between two adjacent resonators is designed to be: $f_{n+1} - f_{n} = \delta  = 2$ MHz
    \item the n-th MKID's resonance frequency is randomly distributed about $ n \delta $ according to a Gaussian distribution $G(f_n,\sigma)= \frac{1}{\sqrt{2\pi}\sigma} e^{{-\frac{1}{2}\left(\frac{(f_n-n\delta)}{\sigma}\right)^2}}$
    \item given a defined window size, a resonator that falls within said distance of any other resonator is said to be clashing.  
\end{enumerate}{}
\begin{figure}
    \centering
    \includegraphics[width=0.75\textwidth]{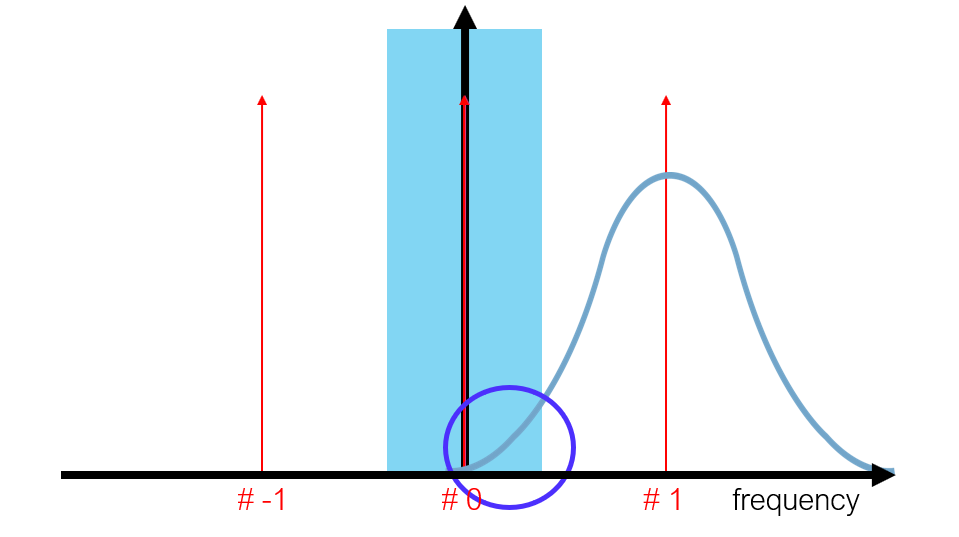}
    \caption{Representation of the simple model of the MKIDs array. Resonances $\# 0$ and $\#\pm1$ are shown as well as the collision window. The probability of two resonators colliding is given by the area underneath the bell curve within the limits of the collision window. }
    \label{model_yield}
\end{figure}{}
 
The probability that the $0$th resonator does not clash with the nth resonator, P$_{n0}$, can be evaluated, according to  Liu et al.\cite{yield_liu} as shown in equation 2, as one minus the overlapping area defined by the gaussian distribution  and the collision window as described in Figure \ref{model_yield}.
Furthermore, due to the translation symmetry of the model, it is only necessary to evaluate P$_{n0}$ for all values of $\pm n$. 
\begin{equation}
    P_{n0} = 1-  \frac{Erf\left(\frac{n\delta + \chi w}{\sqrt{2}\sigma}\right) - Erf\left(\frac{n\delta-\chi w}{\sqrt{2}\sigma}\right)}{2}
\end{equation}{}
Where the integral of a Gaussian is evaluated through the error function Erf(x) defined as in Equation \ref{eq:errorfunction}
\begin{equation}
    Erf(x) = \frac{1}{\sqrt{\pi}}\int\limits_{-x}^{+x} e^{-t^2}dt
    \label{eq:errorfunction}
\end{equation}
Here, $\delta$ is the nominal spacing between the resonators ($\delta$ = 2 MHz is a standard choice \cite{DARKNESS}) , and  $\chi w$  identifies the collision window through two parameters: $w$ represents the resonator's line-width and a $\chi=5$ is assumed under the hypothesis that the resonators are well distinguishable when separated by $5$ line-widths. This is to say that the profile of the $0$-th resonator is well contained in a window five times larges than its typical line-width.\\
Assuming all the resonators to be independent, the total yield is derived as:
\begin{equation}
    Yield = \prod\limits_{n= -n_{max}}^{n_{max}} P_{n0}
    \label{eq:yield}
\end{equation}{}
The assumption of an infinite array allows to identify the probability presented in equation \ref{eq:yield} with the total yield of the array.
\noindent
For the sake of a model, it was assumed  that  the MKID $\#0$ resonates at the central frequency of the working bandwidth, $6$ GHz. And only the first $1000$ resonators left and right have been taken into consideration. Equation \ref{eq:yield} is only valid in the case of an infinite array of evenly spaced resonators.
\\
The analysis of this model will be carried out in the next sections and the dependencies of the yield from physical parameters such as the quality factor of the resonators and the normal state sheet resistance of the superconductor will be investigated. It is worth observing that while increasing the spacing $\Delta$ between two consecutive elements of the array, the yield increases accordingly.\\
\subsection{Effects of the sheet resistivity}
\label{subsec:crit-temp}

Given the change in kinetic inductance, and knowing that the resonance frequency of an LC circuit $f_r = \frac{1}{\sqrt{LC}} $ the fractional frequency shift in resonance frequency induced by a change in sheet resistance R$_s$, therefore a local change in kinetic inductance, is easily obtained  provided three simplifying assumptions are met: 
\begin{enumerate}
    \item The capacitance of the resonator is only due to its geometry
    \item The inductance is only due to the kinetic inductance. The kinetic inductance fraction $\alpha = 1$
    \item Any variation in critical temperature T$_{c0}$ across the wafer is negligible.
\end{enumerate}{}
Under these circumstances, the fractional frequency shift is given by Equation \ref{eq:freq_shift1}\\
\begin{equation}
\label{eq:freq_shift1}
\frac{\Delta f}{f_r}(R_s)= \frac{\Delta R_s}{R_s}
\end{equation}

The main purpose of this calculation is to try and estimate reasonable boundaries for the width $\sigma$ of the distribution of the frequencies about their mean value $f_n$. Given the biggest deviation we have observed in sheet resistance, as described in the previous section, one can conclude that $\frac{\Delta R_s}{R_s} = 0.06 $. It is possible to identify the change in resonance frequency given by such a large variation in sheet resistance, as the value at 3$\sigma$. Such a non uniformity in R$_s$, under these assumptions, yield a value for $\sigma$ of 120 kHZ. \\

\subsection{Fabrication yield - results}
\label{subsec:fab-yieldevaluation}
\begin{figure}[h!]
    \centering
    \includegraphics[width=0.75\textwidth]{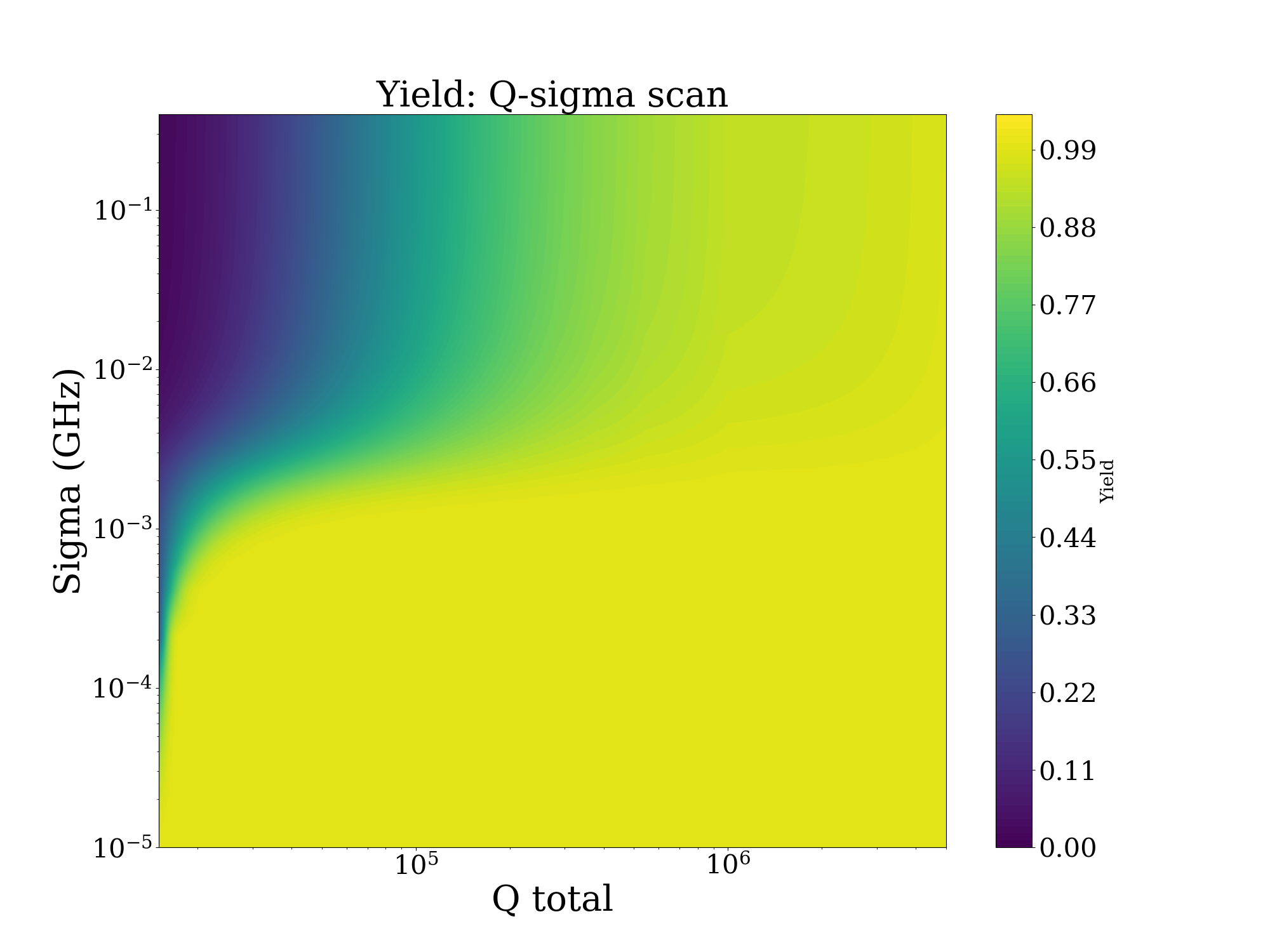}
    \caption{Fabrication yield as a function of the overall quality factor $Q_{tot}$ and the width of the frequency scatter distribution, $\sigma$.}
    \label{yield}
\end{figure}

It is important to discuss the results of the model previously described. The higher total quality factor $Q_{tot}$ of the resonators, the sharper is the resonator's dip. $Q_{tot}$  is  therefore inversely proportional to the line-width $w = \frac{f_0}{Q}$ where $f_0$ is the characteristic resonant frequency.
Assuming a $\delta= 2$ MHz, and based on formula \ref{eq:yield}, Figure \ref{yield} shows a 2D plot of the fabrication yield as a function of both  $\sigma$ and $Q_{tot}$. Two trends are visible in the plot. The yield increases with increasing quality; this says that the sharper the resonator, hence the smaller their line-width and the clashing window, the higher the fabrication yield is. Also, as expected, the smaller the width of the  frequency scatter distribution, the higher the yield per fixed values of $Q_{tot}$. \\
Very strongly coupled resonators with low quality factors, such as $Q_{tot}< 15000$, represent the case where the resonator's rejection window is intrinsically larger than the spacing $\delta$ reducing dramatically the yield to a mathematical zero based on previous assumptions. For very high values of $\sigma$, the model is, in fact, describing an almost random distribution of resonant frequencies. It therefore makes sense that the yield becomes proportional to the rejection window size, thus to the total quality factor $Q_{tot}$.

{}
There are two main issues to address: first of all, a gargantuan value of $\sigma$ would produce a $100\%$ yield as it would scatter most of the resonators outside the working-bandwidth,$4-8$ GHz. It was observed that the yield starts increasing from $\sigma = 1$ GHz onward. Another limit of the model is that, when evaluating the probability of a collision, only one of the two resonator's dips is broadened, while the other is still regarded as if it was infinitely sharp. This effect is only relevant for very strongly coupled resonators, $Q_{tot} < \frac{1}{\sigma}$, therefore it is not  significant for the plot in  Figure \ref{yield}. Further improvement, which is a higher-order approximation to what has been discussed so far resides in the evaluation of the line-width as a function of the specific resonator's resonance frequency. 
In order to summarise the results described in the previous pages, Table \ref{tab:yield} represents some numerical results for interesting values of uniformity ( $\Delta R_s/R_s$) and total quality factor ($Q_{tot}$) .\\
\vspace{0.5cm}
\begin{table}[h!]
    \centering
    \begin{tabular}{|c|c|c|}
    \hline
    $\Delta R_S / R_s$  & $Q_{tot}$ & Yield\,(\%)  \\
    \hline
    
    0.01 & 20000 & 99.9\\
    \hline
    0.06 & 20000 & 94.8\\
    \hline
    0.10 & 20000 & 83.5\\
    \hline
    0.15 & 20000 & 70.7\\
    \hline
    0.01 & 30000 & $\geq99.9$\\
    \hline
    0.06 & 30000 & 99.9\\
    \hline
    0.10 & 30000 & 97.5\\
    \hline
    0.15 & 30000 & 88.6\\
    \hline
    0.01 & 50000 & $\geq99.9$\\
    \hline
    0.06 & 50000 & $\geq99.9$\\
    \hline
    0.10 & 50000 & 97.9\\
    \hline
    0.15 & 50000 & 95.8\\
    \hline
    0.01 & 100000 & $\geq99.9$\\
    \hline
    0.06 & 100000 & $\geq99.9$\\
    \hline
    0.10 & 100000 & 99.6\\
    \hline
    0.15 & 100000 & 98.4\\
    \hline
    \end{tabular}
    \caption{Table representing the yield of a 2000  pixel array as a function of $\Delta R_S / R_s$ and the total quality factor $Q_{tot}$}
    \label{tab:yield}
\end{table}{}
\section{CONCLUSIONS}
In this paper we have characterised TiN/Ti/TiN multilayers with particular interest in their possible application for Microwave Kinetic Inductance Detectors. We demonstrated the possibility of controlling the critical temperature of the multilayer by changing the thickness of the Ti layer. Building on these results, we investigated the uniformity of our superconducting thin films. We have demonstrated that the normal state sheet resistance of the thin film, when measured at room temperature, is uniform within $\pm 6\%$ of its average value. In terms of critical temperature, we have not observed any significant variation within the resolution of our instruments. \\
Finally, we developed a model in order to predict how a $6\%$ variation in resistivity of the thin film affects the pixel yield of an MKIDs array.
We found out that, when no other effects come into play, an array of 2000 resonators evenly spaced by 2 MHz can exhibit a pixel yield of 94.8 $\%$ when the resonators exhibit a total quality factor Q$_{tot}$ of 20000, or more for even higher $Q_{tot}$.\\

\acknowledgments % equivalent to \section*{ACKNOWLEDGMENTS}       
 
This publication has emanated from research conducted with the financial support of Science Foundation Ireland under Grant number 15/IA/2880. We would like to also thank the Process \& Product Development team at Tyndall National Institute for providing wafer-processing services to fabricate the devices

% References
\bibliography{report} % bibliography data in report.bib
\bibliographystyle{spiebib} % makes bibtex use spiebib.bst

\end{document}